\documentclass{article}
\usepackage{PRIMEarxiv}

\usepackage[utf8]{inputenc} 
\usepackage[T1]{fontenc}    
\usepackage{hyperref}       
\usepackage{url}            
\usepackage{booktabs}       
\usepackage{amsfonts}       
\usepackage{nicefrac}       
\usepackage{microtype}      
\usepackage{lipsum}
\usepackage{fancyhdr}       
\usepackage{graphicx}       
\graphicspath{{media/}}     

\usepackage[T1]{fontenc}
\usepackage{textcomp}
\usepackage{amsmath}
\usepackage{amssymb}
\usepackage{soul}
\usepackage{bm}
\usepackage{mdwmath}
\usepackage{cases}
\usepackage{graphicx}
\graphicspath{{Figure/PDF/}{Biography/PDF/}}
\usepackage[dvipsnames]{xcolor}
\definecolor{myblue}{rgb}{0,0.4980,1} 
\definecolor{myred}{rgb}{0.8706,0.1608,0.0627} 
\usepackage[labelformat=simple]{subcaption}

\usepackage[nosort,noadjust]{cite}
\usepackage[group-separator= {,},range-units=single,range-phrase=~--~]{siunitx}
\usepackage{url}
\usepackage{tabularray}
\UseTblrLibrary{counter}
\UseTblrLibrary{diagbox}

\usepackage{hyperref}
\newcommand{\colorhypersetup}{\@ifpackageloaded{hyperref}{\hypersetup{%
	bookmarksopen=true,%
	bookmarksnumbered=true,%
	pdfpagemode={UseOutlines},
	pdfstartview={FitH},%
	colorlinks=true,%
	linkcolor={myred},%
	citecolor={orange}
}}{\empty}}
\newcommand{\blackhypersetup}{\@ifpackageloaded{hyperref}{\hypersetup{%
	bookmarksopen=true,%
	bookmarksnumbered=true,%
	pdfpagemode={UseOutlines},
	pdfstartview={FitH},%
	colorlinks=true,%
	allcolors={black}
}}{\empty}}

\usepackage{mfirstuc-english}
\MFUhyphentrue
\usepackage[version3,description,IEEEtran]{eacro}
\acsetup{%
    pages/display=all,%
    pages/seq/use=false,%
    pages/name=true,%
    pages/fill={\quad},%
    make-links=true%
}
\DeclareAcronym{iot}{
	short = IoT,
	long = Internet of things}
\DeclareAcronym{ai}{
    short = AI,
    long = artificial intelligence}
\DeclareAcronym{2g}{
    short = 2G,
    long = second generation}
\DeclareAcronym{3g}{
    short = 3G,
    long = third generation}
\DeclareAcronym{4g}{
    short = 4G,
    long = fourth generation}
\DeclareAcronym{5g}{
    short = 5G,
    long = fifth generation}
\DeclareAcronym{6g}{
    short = 6G,
    long = sixth generation}
\DeclareAcronym{nbiot}{
    short = NB-IoT,
    long = narrowband \acs*{iot}}
\DeclareAcronym{3gpp}{
    short = 3GPP,
    long = 3rd Generation Partnership Project}
\DeclareAcronym{bs}{
    short = BS,
    long = base station}
\DeclareAcronym{ue}{
    short = UE,
    long = user equipment}
\DeclareAcronym{csi}{
    short = CSI,
    long = channel state information}
\DeclareAcronym{snr}{
    short = SNR,
    long = signal-to-interference ratio}
\DeclareAcronym{iov}{
    short = IoV,
    long =  Internet of Vehicle}
\DeclareAcronym{vdtn}{
    short = VDTN,
    long =  Vehicular Digital Twin Network}
\DeclareAcronym{rra}{
    short = RRA,
    long =  Radio Resource Allocation}
\DeclareAcronym{drl}{
    short = DRL,
    long = Deep Reinforcement Learning}
\DeclareAcronym{voi}{
    short = VoI,
    long = Value of Information}

\usepackage{algpseudocode}
\newcounter{MYalgorithmic}

{%
	\vspace*{3pt}
	\hrule height 1pt}
\newcounter{MYitem}[MYalgorithmic]

\newcommand{\MYlabel}[1]{\def\@currentlabel{\theALG@line}\label{#1}}

\newcommand{\upperroman}[1]{\uppercase\expandafter{\romannumeral#1}}

\newcommand{\myunit}[1]{%
	\ifmmode
		\mathrm{#1}
	\else
		$ \mathrm{#1} $
	\fi}
\newcommand{\murm}{%
	\ifmmode
		\text{\textmu}
	\else
		\textmu
	\fi}

\newlength{\mysinglefigwidth}
\newlength{\mymultifigwidth}
\AtBeginDocument{\setlength{\mysinglefigwidth}{0.8\linewidth}}
\AtBeginDocument{\setlength{\mymultifigwidth}{0.5\linewidth}}

\usepackage{enumitem}

\makeatother

\begin{document}

\title{An SDR-Based Test Platform for 5G NTN Prototyping and Validation
\thanks{\textit{\underline{Citation}}:
\textbf{L. Hou, K. Zheng, J. Mei and C. Huang, "An SDR-Based Test Platform for 5G NTN Prototyping and Validation," in IEEE Open Journal of the Communications Society, doi: 10.1109/OJCOMS.2025.3610869.}}
}

\author{
  Lu Hou \\
  Beijing University of Posts and Telecommunications \\
  Beijing\\
  \texttt{houlu8674@bupt.edu.cn} \\
   \And
  Kan Zheng, Jie Mei \\
Ningbo University\\
  Ningbo \\
  \texttt{zhengkan,meijie@nbu.edu.cn} \\
  \And
  Cheng Huang\\
  Geeflex Ltd.\\
  Beijing\\
  \texttt{cheng.huang@geeflex.com}
}

\maketitle

\begin{abstract}
  
The integration of satellite communication into 5G has been formalized in 3GPP Release 17 through the specification of Non-Terrestrial Networks (NTN), marking a significant step toward achieving global connectivity. However, the early-stage maturity of 5G NTN standards and the lack of commercial NTN-capable equipment hinder extensive performance validation and system prototyping. To address this gap, this paper proposes a software-defined radio (SDR) test platform with General-Purpose Processor (GPP) processing, leveraging Amarisoft’s 5G NTN protocol stack software while performing custom system integration and adaptation for real satellite operation. The platform supports bidirectional communication between an SDR-based NTN gNB and UE emulator through a Geostationary Earth Orbit (GEO) satellite link, with full compliance to 3GPP NTN specifications. We provide detailed insights into the system architecture, SDR hardware–software co-design, and satellite gateway adaptations. Through field trials, we evaluate the performance metrics including downlink throughput and round-trip time. Results validate the feasibility and effectiveness of SDR-based platforms for NTN testing, and highlight their potential in bridging current implementation gaps before widespread commercial deployment.

\end{abstract}


\maketitle

\section{Introduction}
\label{sec:Introduction}

\acresetall

The integration of satellite technology into 5G networks has been actively pursued in recent years, achieving significant milestones in recent years~\cite{TR_38_811,TR_38_821}. With the inclusion of Non-Terrestrial Network (NTN) specifications in the $3^{\mathrm{rd}}$ Generation Partnership Project (3GPP), 5G systems have officially expanded their scope to encompass satellite communications~\cite{NTNSurvey}. This marks a critical step toward ubiquitous connectivity, bridging terrestrial and non-terrestrial networks. As 5G evolves toward 6G, NTN is expected to play an essential role in enabling global coverage, enhanced reliability, and seamless mobility—particularly for remote, maritime, and aerial use cases~\cite{Nguyen2024,6GNTN}. 
\par

Given that 5G NTN specifications are still in their early stages, commercial gNodeBs (gNBs) and User Equipments (UEs) with full NTN support are not yet widely available. Nevertheless, the industry and academia have an urgent need to conduct functional validation and performance testing in real-world scenarios~\cite{Muro2024}. Software-Defined Radio (SDR) technology emerges as a compelling solution for bridging this gap due to its inherent flexibility, reconfigurability, and cost-efficiency. SDR-based platforms enable researchers and engineers to implement NTN features in software, facilitating early testing and iterative refinement as 3GPP standards evolve.
\par

In practice, most SDR solutions are host-based, where baseband signal processing is performed on General-Purpose Processors (GPPs) rather than on dedicated accelerators such as FPGAs or DSPs~\cite{GPPSDR}. Within this paradigm, software frameworks such as OpenAirInterface (OAI), srsRAN, and Amarisoft have become widely adopted in the research and testing of 4G and 5G systems~\cite{OAINTN,OAI2,OAITEST,SRSRANOATCOMP}. Coupled with SDR hardware front-ends, these software suites have played a pivotal role in both academia and industry by enabling flexible experimentation, rapid prototyping, and technology validation, thereby accelerating the evolution and standardization of cellular networks.

Among these frameworks, OAI stand out as open-source software that provides 3GPP-compliant implementations of 4G and 5G cellular network technologies~\cite{OAIWebsite}. Their openness and customizability make them well suited for algorithm design, protocol testing, and proof-of-concept demonstrations. Such characteristics explain their prominence in academic research communities. However, the current NTN functionality in OAI remains limited~\cite{Kumar2024}. Although the OAI community has introduced an initial 5G NTN framework compliant with Release 17 and demonstrated it through laboratory experiments and over-the-air testing, the implementation does not yet deliver the complete NTN feature set. For example, full support of System Information Block 19 (SIB-19)—a critical function for NTN operation—is still missing, which forces the access and synchronization procedures to depend on extensive manual configuration and tuning. More broadly, the OAI platform still relies heavily on manual intervention, lacks automatic gain control (AGC), and employs synchronization mechanisms that are not sufficiently robust, thereby limiting its suitability for automated deployment in practical scenarios. While OAI researchers have explored various workarounds to mitigate specific challenges, such as those encountered in GEO satellite testing~\cite{Kumar2025}, these efforts do not fundamentally compensate for the missing Release-17 functionalities.

By contrast, Amarisoft delivers a commercial-grade software suite that fully supports both 3GPP Releases 16 and 17, offering comprehensive coverage of NTN-specific functionalities and thereby enabling more complete end-to-end validation.
 
\par

To this end, this paper presents a GPP-based SDR test platform for 5G NTN, which builds upon the commercial-grade software stack provided by Amarisoft. Rather than directly relying on off-the-shelf software, the platform achieves innovation through customized system integration and parameter adaptation, jointly optimized with Amarisoft to support transparent GEO satellite operation. The proposed system strictly follows 3GPP specifications and establishes a bidirectional, end-to-end communication link between an SDR-based NTN gNB and an NTN UE emulator over actual GEO satellite channels. To the best of our knowledge, this represents the first 3GPP-compliant GEO satellite NTN test platform validated through real-world experimentation. The system incorporates three essential components: (1) optimized SDR devices with hardware–software co-design for real-time signal processing; (2) a frequency-conversion satellite gateway bridging sub-6 GHz SDR hardware with Ku-band satellite operation; and (3) a complete 5G protocol stack implementation enhanced with NTN-specific extensions.

The main contributions of this paper are summarized as follows, i.e., 
\begin{itemize}
\item We develop a flexible SDR-based architecture with GPP processing that leverages Amarisoft’s 5G NTN protocol stack as the foundation. Unlike the standalone software, our design emphasizes customized system integration, hardware–software co-design, and parameter tuning to accommodate transparent GEO satellite channels.
\item A fully functional SDR test platform is designed and implemented to enable 3GPP-compliant 5G NTN testing, supporting a bidirectional end-to-end link between an SDR-based NTN gNB and an NTN UE emulator through a real GEO satellite relay.
\item Comprehensive field trials are conducted over a commercial GEO satellite, during which critical parameters are jointly adapted and validated in collaboration with Amarisoft. The experimental results, including throughput and round-trip latency, demonstrate the feasibility and effectiveness of the platform for practical 5G NTN prototyping and validation.
\end{itemize}

The remainder of this paper is organized as follows: Section~\ref{sec:NTN} provides a brief overview of the 5G NTN. Section~\ref{sec:sdr} describes the proposed SDR-based testing platform for 5G NTN in detail. Section~\ref{sec:Results} presents experimental results obtained through field trials. Finally, conclusions and future directions are summarized in Section~\ref{sec:Conclusion}.

\section{Overview of 5G NTN}
\label{sec:NTN}

To contextualize the proposed SDR-based 5G NTN test platform, this section first outlines the ongoing 3GPP standardization progress of 5G NTN, followed by the fundamental architecture of satellite-based NTN systems.

\subsection{Standardization in 3GPP}


3GPP has been engaged in the discussion and standardization of NTN technologies since 2019. Beginning with Release 17, 3GPP initiated the formal specification development for NTN integration into the existing 5G ecosystem. As part of Release 17, NTN comprises New Radio (NR) NTN and Internet-of-Things (IoT) NTN, aiming to facilitate the rapid deployment of satellite services in 5G. Mobile devices and IoT terminals are capable of directly connecting to satellites and can access mobile cellular networks using either 5G NR or NB-IoT/eMTC protocols. In parallel, several enhanced technologies for both NR NTN and IoT NTN systems have been explored, covering aspects such as network architecture, time-frequency synchronization, and mobility. Meanwhile, IoT NTN addresses specific requirements for non-continuous coverage. In Release 18, 3GPP continues to study a variety of new technologies to further enhance the satellite networking capabilities of 5G NTN, including support for deployment in frequency bands above \qty{10}{\giga\hertz}, improved coverage, enhanced mobility, and service continuity~\cite{NTNR17R18}.
\par


\subsection{Satellite-based 5G NTN Architecture}

The 5G NTN family includes satellite, high altitude platform systems (HAPS), and air-to-ground networks. Here we only focus on satellite-based NTN networks, which mainly consists of satellite, gateway and UEs, i.e., 
\begin{itemize}
\item \textbf{Satellite}:
The satellite functions as a relay station, providing connectivity between the gateway and UEs. Based on orbital altitude, satellites are generally categorized into GEO, Medium Earth Orbit (MEO), and Low Earth Orbit (LEO). To enhance system capacity, satellites typically utilize Radio Frequency (RF) signals with single or multiple beams over their service areas. The coverage area of each beam is determined by the satellite's antenna configuration and elevation angle.

\item \textbf{NTN Gateway}: 
The NTN gateway acts as the bridge for UEs in 5G NTN to access the 5G Core Network (CN). A GEO satellite maintains connections with one or more NTN gateways located within its signal coverage. In such systems, UEs in the service area are typically served by a specific and single gateway. In contrast, non-GEO satellites such as LEO and MEO, due to their high-speed movement relative to the gateway, need to be served sequentially by multiple gateways distributed along their orbital paths. The 5G NTN system must ensure link availability and continuity between adjacent serving gateways, with sufficient time for mobility anchoring and handover.

\item \textbf{UE}:
UEs in 5G NTN systems include a diverse range of terminals such as handheld, vehicular, maritime, airborne, and embedded devices designed for IoT applications. These UEs are capable of establishing communication with satellites within the defined service area, thereby enabling seamless connectivity and data exchange.
 
\end{itemize}
 
In 5G NTN systems, two essential links facilitate communication, i.e., the service link connecting UEs with the satellite, and the feeder link between the NTN gateway to the satellite. Additionally, in satellite constellations, Inter-Satellite Links (ISLs) are optional connections requiring regenerative payloads onboard the satellites. ISLs, operating in either RF or optical bands, can improve overall network efficiency. These links allow satellites within a constellation to relay signals, thereby optimizing network performance and adaptability.

\begin{figure}[!t]
    \subfloat[Transparent satellite-based RAN.]{\includegraphics[width = 0.95\mymultifigwidth]{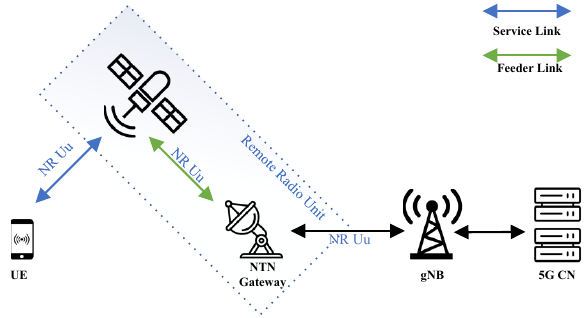}\label{fig_RAN1}}
    \hfil
    \subfloat[Regenerative satellite-based RANs.]{\includegraphics[width = 0.95\mymultifigwidth]{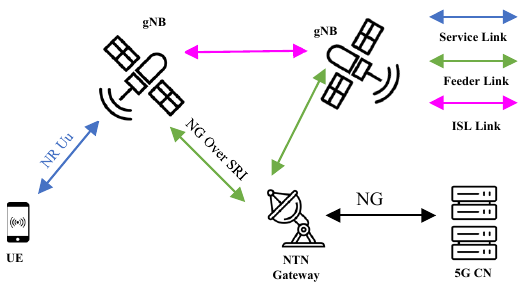}\label{fig_RAN2}}
    \caption{5G NTN-based RAN architectures.}
    \label{fig_RAN}
\end{figure}

As illustrated in Fig.~\ref{fig_RAN}, 5G NTN systems have two typical Radio Access Network (RAN) architectures, determined by the type of satellite payload, i.e., 
\begin{itemize}
\item \textbf{Transparent satellite-based RAN}:
The satellite payload performs only RF signal filtering, frequency conversion, and amplification in both uplink and downlink directions, leaving the waveform signal unchanged. In essence, it functions as an analog RF repeater, repeating the NR Uu radio interface from the feeder link to the service link and vice versa. Accordingly, the Satellite Radio Interface (SRI) on the feeder link is the NR Uu. The NTN gateway incorporates all functionalities required to process and forward NR-Uu signals. Moreover, multiple transparent satellites can be connected to a single gNB on the ground.
\par
This architecture is suitable for both new satellite technologies and the reuse of existing satellite resources (typically equipped with transparent payload). It facilitates the rapid commercial deployment of 5G NTN systems.

\item  \textbf{Regenerative satellite-based RAN}:
The satellite payload undertakes signal regeneration received from UEs, involving baseband signal processing such as modulation and coding. This is to implement all or part of base station functions onboard the satellite. In this case, the NR-Uu interface exists only on the service link. Furthermore, the satellite payload supports Inter-Satellite Links (ISL) between satellites. The NTN gateway operates as a transport network layer node, supporting all essential transport protocols. The gNBs onboard different satellites may connect to the same 5G CN on the ground. If a satellite hosts multiple gNBs, the same SRI transports all corresponding Next Generation (NG) interface instances.
\par

This architecture features flexible networking, low transmission delay, and the ability to flexibly schedule hopping beam resources. However, it comes with heightened technical complexity and requires the launch of new satellites, leading to increased implementation costs for the 5G NTN systems.

\end{itemize}

It is necessary to note that current 5G NTN systems, as defined in both Release 17 and Release 18, operate in transparent communication mode, wherein the satellite merely forwards RF signals without performing baseband processing. In the upcoming Release 19, 3GPP will focus on studying NTN systems operating in regenerative mode~\cite{Zhang2023}.

Given the significance and implementation complexity of NR NTN, it is designated as the primary focus of our testing platform. Accordingly, unless otherwise specified, the term ``NTN'' hereafter refers specifically to NR NTN throughout the remainder of this paper.
 
\section{SDR-based platform for 5G NTN testing}
\label{sec:sdr}

 \begin{figure*}[!t]
	\centering
	\includegraphics[width=1.0\linewidth]{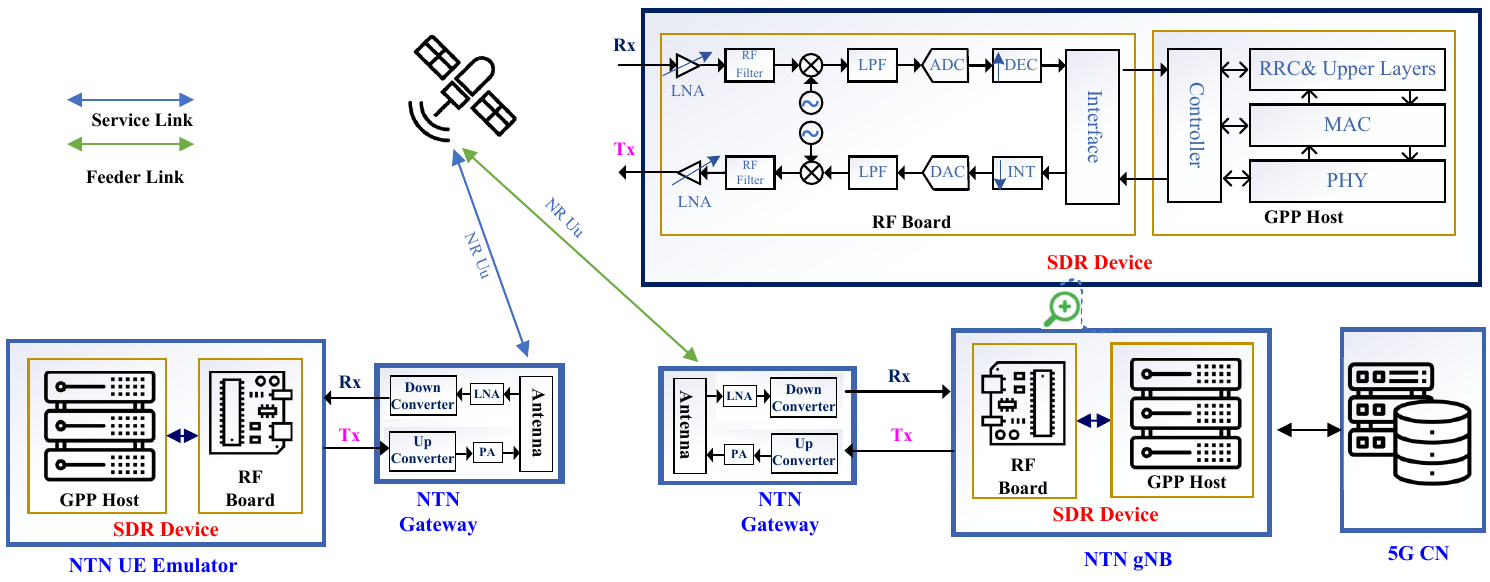}
	\caption{Illustration of GPP-based SDR platform for 5G NTN testing.}
	\label{fig_system}
\end{figure*}

As illustrated in Fig.~\ref{fig_system}, we propose a GPP-based SDR platform designed to comprehensively evaluate the functional and performance characteristics of 5G NTN systems with transparent satellite-based RAN. The platform supports end-to-end testing scenarios under both controlled laboratory and real-world field conditions, assuming a transparent payload satellite for signal relaying to enable precise evaluation of 5G NTN systems.

Due to the unavailability of commercial 5G NTN-capable UEs, our platform employs a dual-SDR architecture where one SDR device functions as a gNB and another as a UE emulator. Both devices share identical hardware, with roles determined by software implementation, highlighting the inherent flexibility of SDR-based testing platforms.

Most commercially available SDR devices operate in sub-6 \unit{\giga\hertz} bands, whereas satellite systems often use higher frequencies (e.g., Ku or Ka bands). To bridge this gap, our platform incorporates an NTN gateway with a frequency conversion module, enabling seamless translation between SDR signals and satellite bands. SDR devices connect to the gateway via RF cables carrying sub-6 \unit{\giga\hertz} signals.

In our architecture, the 5G CN interfaces with the SDR-based NTN gNB via the standard NG interface. The gNB connects to the NTN gateway that establishes the feeder link to a transparent satellite, which then communicates with the NTN UE emulator via the NR Uu interface, completing the end-to-end NTN connection.

\subsection{SDR Device}

The SDR device is the core of our testing platform, adopting a GPP-based solution with Amarisoft software, chosen for flexibility and reliability in mobile network testing. As shown in Fig.~\ref{fig_system}, it consists of two main subsystems: (1) an RF board for RF signal generation, conversion, and transmission, and (2) a GPP host that executes baseband processing and protocol stacks.

A high-speed data interface (PCIe) enables real-time bidirectional exchange of I/Q samples, providing both high throughput (exceeding \qty{8}{\giga T\per\second}) and low latency, essential for 5G signal processing.

\subsubsection{RF board:} 

Our RF board employs a zero-IF architecture, eliminating intermediate frequency stages to reduce complexity and improve power efficiency. It performs bidirectional RF–baseband conversion and supports full-duplex operation.

On the receive path, the RF signal is amplified, filtered, downconverted using a digitally tunable local oscillator, and passed through a low-pass filter. The resulting analog signal is digitized and decimated to produce baseband digital samples, which are transferred to the GPP host via a high-speed interface. On the transmit path, baseband samples are interpolated, converted to analog via DAC, upconverted to the target RF frequency, filtered, amplified, and transmitted. 
 
To facilitate efficient data movement, our RF board incorporates dedicated Direct Memory Access (DMA) engines. These hardware-accelerated modules enable direct, high-throughput data transfers between the RF front-end and the memory of the GPP host without CPU intervention, thereby minimizing latency and maximizing data throughput.

\subsubsection{GPP host:}

The GPP host serves as the computational backbone of the SDR platform, executing the full 5G protocol stack—including the Physical (PHY), Medium Access Control (MAC), and higher-layer functionalities—for both the gNB and UE emulator configurations in 5G NTN systems. While GPP-based SDR solutions offer exceptional programming flexibility and adaptability, the computational demands of real-time wireless signal processing, particularly in NTN scenarios characterized by extended propagation delays and pronounced Doppler effects, present significant challenges in achieving deterministic performance and system stability.
\par
To overcome these challenges, the platform integrates a suite of hardware and software optimizations, i.e.,

\begin{itemize}
\item \textbf{Hardware}

The proposed SDR implementation leverages commercial off-the-shelf (COTS) hardware based on the x86 architecture, tailored to meet the processing demands of 5G NTN systems. At its core, a high-performance multi-core CPU enables parallel execution across PHY-layer signal processing chains, MAC-layer scheduling algorithms, and upper-layer protocol stack functions. To complement this, the SDR device adopts a channel-based memory allocation strategy, in which each cell instance is assigned a dedicated memory channel. Accordingly, a minimum dual-channel DDR4 configuration is recommended to ensure reliable baseline performance.

For the implementations of 5G NTN systems operating at typical bandwidths of \qty{5}{\mega\hertz}, our results indicate that this baseline configuration, i.e., a \qty{3.0}{\giga\hertz} quad-core CPU with dual-channel memory, is sufficient to meet processing requirements. The relatively narrow bandwidth of current NTN deployments substantially reduces the baseband processing load compared to terrestrial 5G NR systems, enabling efficient operation while maintaining full compliance with 3GPP protocols.

\item \textbf{Software}

The SDR software comprises three fundamental components, i.e., a) Vectorized Instruction Set for High-Efficiency Processing:  
Leveraging the x86 architecture with AVX2 extensions, the platform exploits parallel vector processing to accelerate computationally intensive operations and maximize throughput.
b) Kernel Optimizations with Adaptive Resource Management:
Custom-developed drivers handle high-throughput I/Q data streams from the RF board using DMA, which is particularly critical in multi-antenna scenarios. The system further implements dynamic resource management that continuously balances CPU cores, memory, and RF interfaces, thereby safeguarding real-time performance and avoiding packet loss when background processes contend for resources.
c) Protocol Stack with NTN-Specific Adaptations:
At the functional core, optimized C implementations of PHY, MAC, and upper-layer protocols employ dynamic multi-threading that scales automatically with processing load. The complete stack integrates NTN-specific features, including extended timing advance and modified Hybrid Automatic Repeat reQuest (HARQ), ensuring compatibility with satellite communication requirements.

\end{itemize}

The hardware–software co-design strategies enable the platform to meet the stringent timing and reliability requirements of 5G NTN, while preserving the flexibility and reconfigurability that define SDR architectures.

\subsection{NTN Gateway}

The ground-based NTN gateway serves as the essential terrestrial interface in our testing platform, enabling reliable satellite communication in conjunction with a SDR device. Strategically deployed at the Earth's surface, the NTN gateway is engineered to provide the necessary RF power and sensitivity for satellite access, leveraging a rigorously designed RF front-end that includes high-power amplifiers, low-noise receiver chains, and precision frequency stabilization to satisfy stringent link budget requirements.
\par
As shown in Fig.~\ref{fig_system}, to interoperate with the SDR device, which is typically constrained to sub-6 \unit{\giga\hertz} operation, the NTN gateway integrates a frequency conversion subsystem that enables seamless compatibility. On the receive side, it downconverts satellite signals to frequencies within the SDR's supported range. On the transmit side, the SDR's output signal (e.g., S-band at \qty{2}{\giga\hertz}) is upconverted to the satellite uplink band (e.g., Ku-band at \qty{14}{\giga\hertz}) and amplified by a Power Amplifier (PA) before transmission.
\par

\subsection{5G CN}

The 5G CN is comparatively lightweight and demands significantly fewer computational resources than the RAN. As a result, it can either be deployed independently on a general-purpose computer or co-located with the SDR-based NTN gNB on the same GPP host. In principle, any 5G CN implementing the standard interface is test-compatible, including open-source options like OAI CN or Open5GS, as well as commercial solutions~\cite{5GCNCOM}.

In summary, the architecture enables systematic validation of NTN protocols within practical SDR constraints, providing researchers with a flexible and accurate means to evaluate performance under real satellite link conditions. Its modular gateway design further facilitates seamless adaptation to diverse satellite frequency bands and regulatory requirements.

\section{Experimental Results and Analysis}
\label{sec:Results}

To verify the feasibility and effectiveness of the SDR-based NTN testing system proposed in this paper, we conducted a series of field experiments using the communication payload of the GEO satellite, i.e., AsiaSat9~\cite{asiasat9}.

\subsection{Experimental Scenarios} 


In our experiments, a high-performance commercial PC with a 16-core AMD processor running at \qty{4.5}{\giga\hertz} and \qty{32}{\giga\byte} of memory serves as the GPP host for the SDR device. A self-developed RF front-end board, FlexSDR400\footnote{https://www.geeflex.com/flexsdr/flexsdr400.html}, is integrated into the SDR device. It supports bandwidths from \qty{5}{\mega\hertz} to \qty{1000}{\mega\hertz} and carrier frequencies up to \qty{7}{\giga\hertz}. Thanks to its zero-IF architecture, the board directly downconverts RF signals to baseband, reducing hardware complexity while maintaining high-fidelity signal processing. Its multi-channel, high-linearity design makes it particularly suitable for SDR-based 5G NTN prototyping and GEO satellite validation. Each SDR device is connected to a commercial NTN gateway equipped with a dish antenna.


\begin{figure}[!t]
    \centering
    \includegraphics[width=0.8\textwidth]{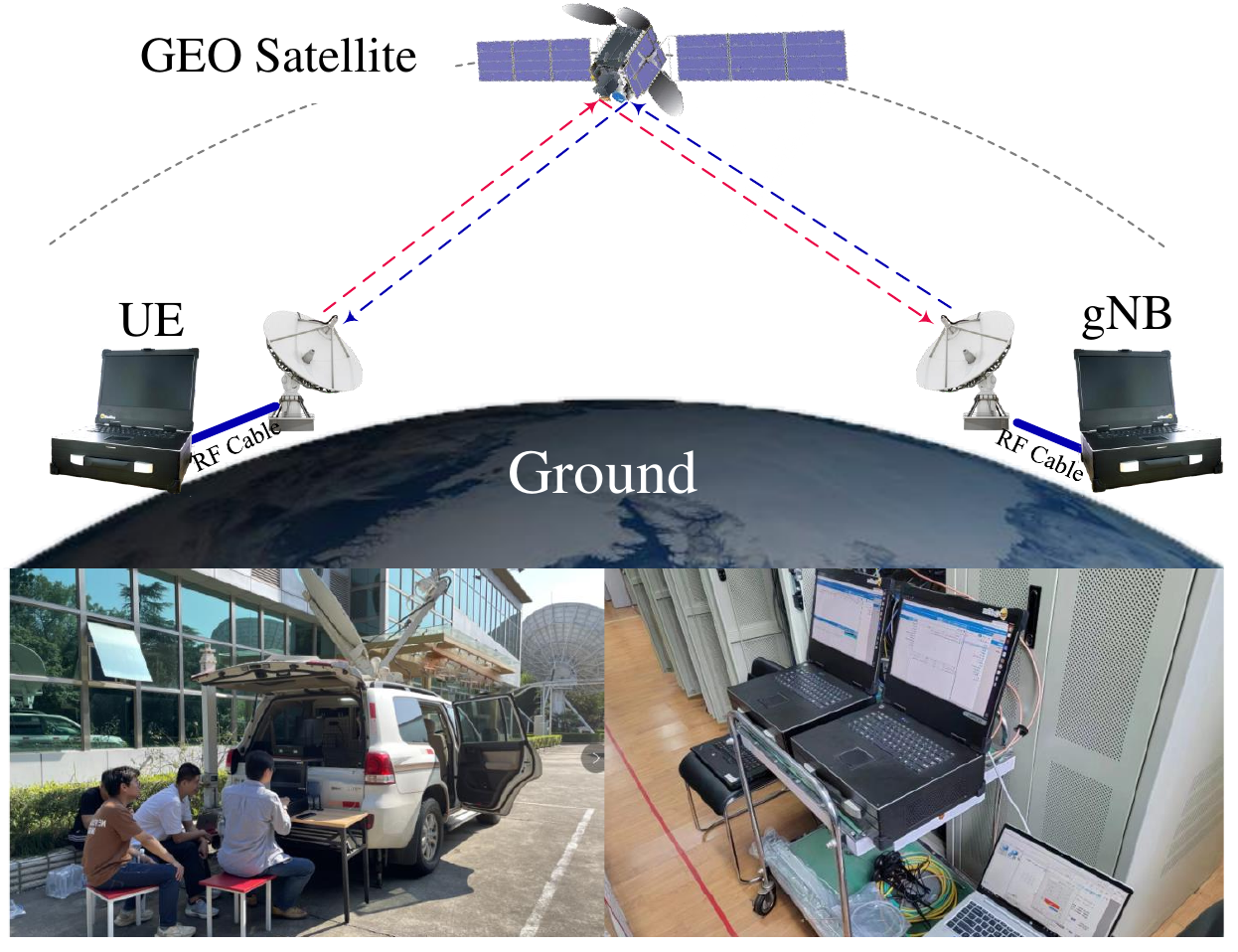}
    \caption{Illustration of the experimental scenarios of the field trials.}
    \label{fig_sim}
\end{figure}

\begin{figure}[!t]
    \centering
    \includegraphics[width=0.8\textwidth]{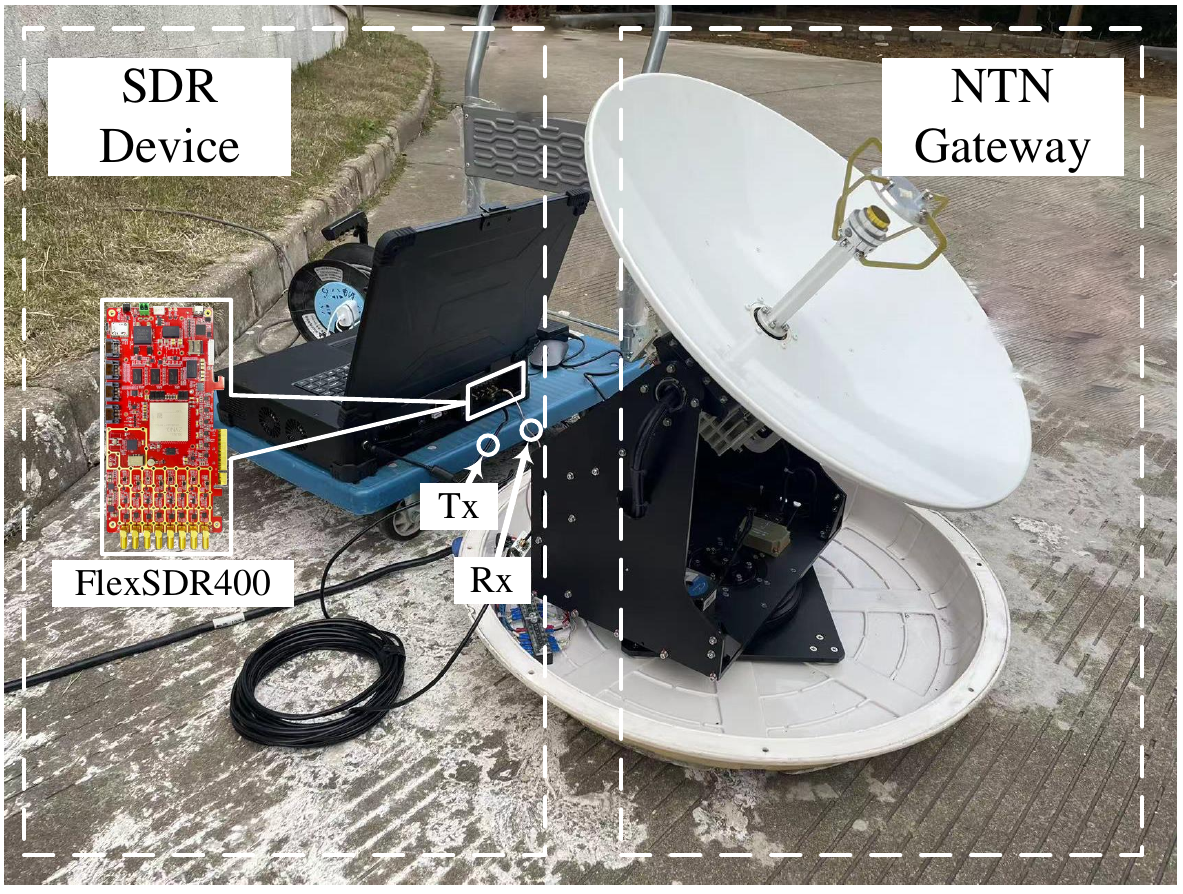}
    \caption{Illustration of UE emulator implemented on the SDR device.}
    \label{fig_device}
\end{figure}

As shown in Fig.~\ref{fig_sim}, two SDR devices were deployed in the field trials. One is configured to function as the NTN gNB, with the 5G CN co-located on the same device. This integrated unit is installed at the satellite ground station, with its transmit and receive ports connected to the corresponding input and output ports of the NTN gateway via RF cables. To establish a bidirectional communication link, the second SDR device is configured to emulate NTN UE functionality, as illustrated in Fig.~\ref{fig_device}. This UE emulator is deployed at a remote location approximately \qty{1060}{\kilo\meter} away from the NTN gNB. For portability, the UE emulator utilizes an NTN gateway equipped with a compact satellite antenna.
\par

To successfully establish the link, several parameters on both the gNB and UE—such as PRACH configurations, geographical positions, and RF frequencies—were carefully adjusted. During this setup process, we observed an end-to-end frequency shift of approximately 3–6 kHz, caused by multiple frequency conversion stages, which can potentially lead to access failures. Although manually adjusting the UE transmission frequency could temporarily restore access, an new argument called ``large\_freq\_shift'' is implemented in the SDR-based gNB, which enables calibration of large and unpredictable frequency errors during PRACH reception.
\par

Both the SDR-based gNB and the UE emulator operate with a \qty{5}{\mega\hertz} channel bandwidth in Frequency Division Duplexing (FDD) mode within the n256 band, allocating \qtyrange{1980}{2010}{\mega\hertz} for uplink and \qtyrange{2170}{2200}{\mega\hertz} for downlink. Since only GEO satellite payloads operating in the Ku band are available in our experiments, both the signals from the gNB and UE emulator have to be converted between the n256 band and the Ku band at the NTN gateway in order to establish satellite connectivity. The main experimental parameters are summarized in Table~\ref{table:experiment_configuration}. 
\par

\newcounter{mycnta}
\newcommand{\mycnta}{\stepcounter{mycnta}(\alph{mycnta})}
\begin{table}[!t]
\centering
	\centering
	\caption{Main parameters of the testing system} 
	\begin{tabular}{|l|c|}
		\hline
		  \textbf{Parameter} & \textbf{Value} \\
		\hline
		  EIRP of satellite gateway at NTN gNB side & $75$ \myunit{dBW} \\
		\hline
            G/T of satellite gateway at NTN gNB side & $30$ \myunit{dB/K} \\
            \hline
            EIRP of GEO satellite & $54$ \myunit{dBW} \\
		\hline
            G/T of GEO satellite & $\geq 6.5$ \myunit{dB/K} \\
		\hline
            EIRP of satellite gateway at NTN UE side & $55$ \myunit{dBW} \\
		\hline
            G/T of satellite gateway at NTN UE side & $16.5$ \myunit{dB/K} \\
		\hline
            Frequency band of service link & Ku \\
		\hline
            Frequency band of feeder link & Ku \\
            \hline
            Frequency band of NR NTN   & n256 \\
		\hline
            Frequency bandwidth of NR NTN  & $5$ \myunit{MHz} \\
		\hline
	\end{tabular}
	\label{table:experiment_configuration}
\end{table}

\begin{figure}[!t]
    \centering
    \includegraphics[width=0.8\textwidth]{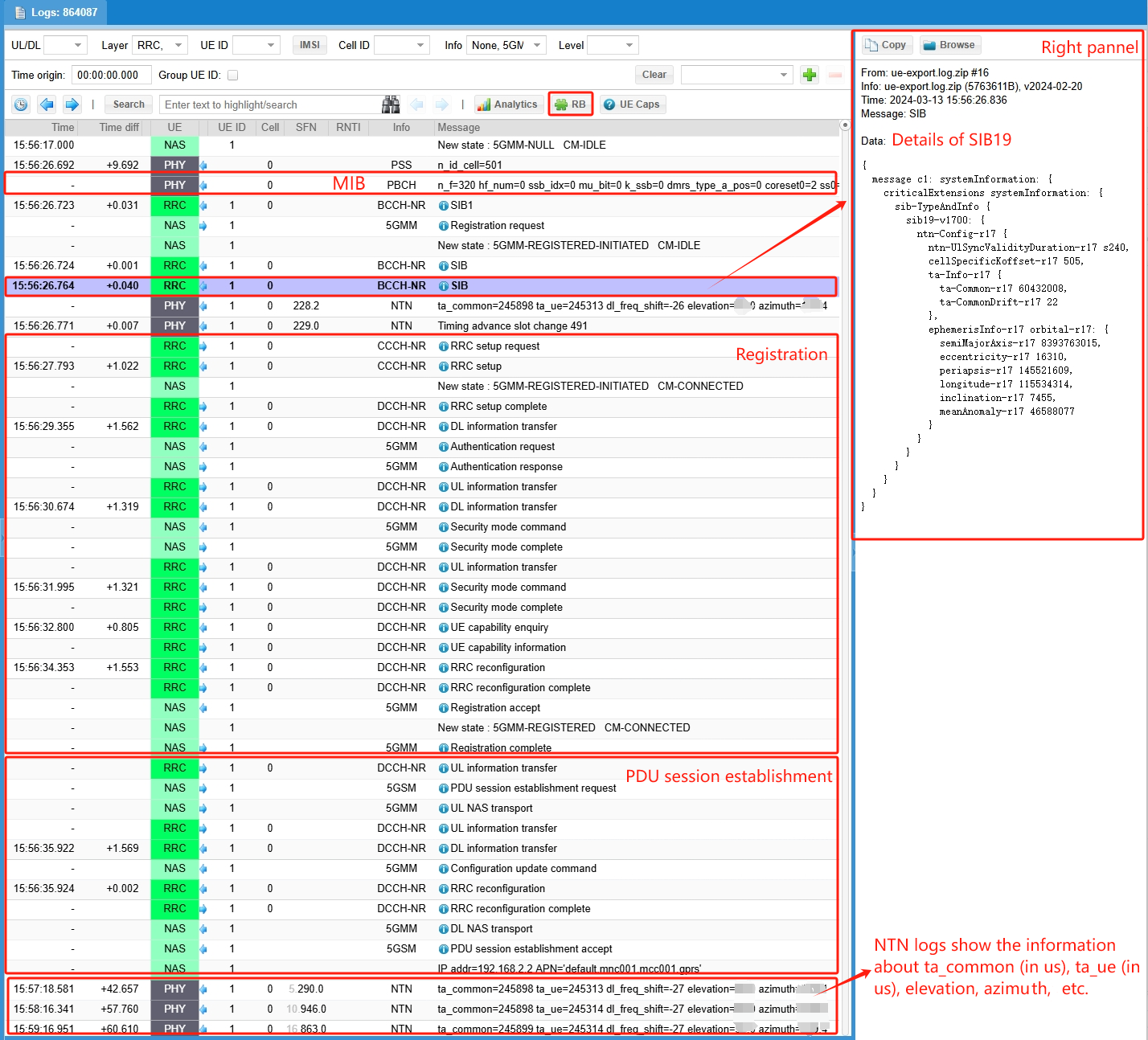}
    \caption{Example of UE log captured during testing}
    \label{fig_uelog}
\end{figure}
\begin{figure}[!t]
    \centering
    \includegraphics[width=0.8\textwidth]{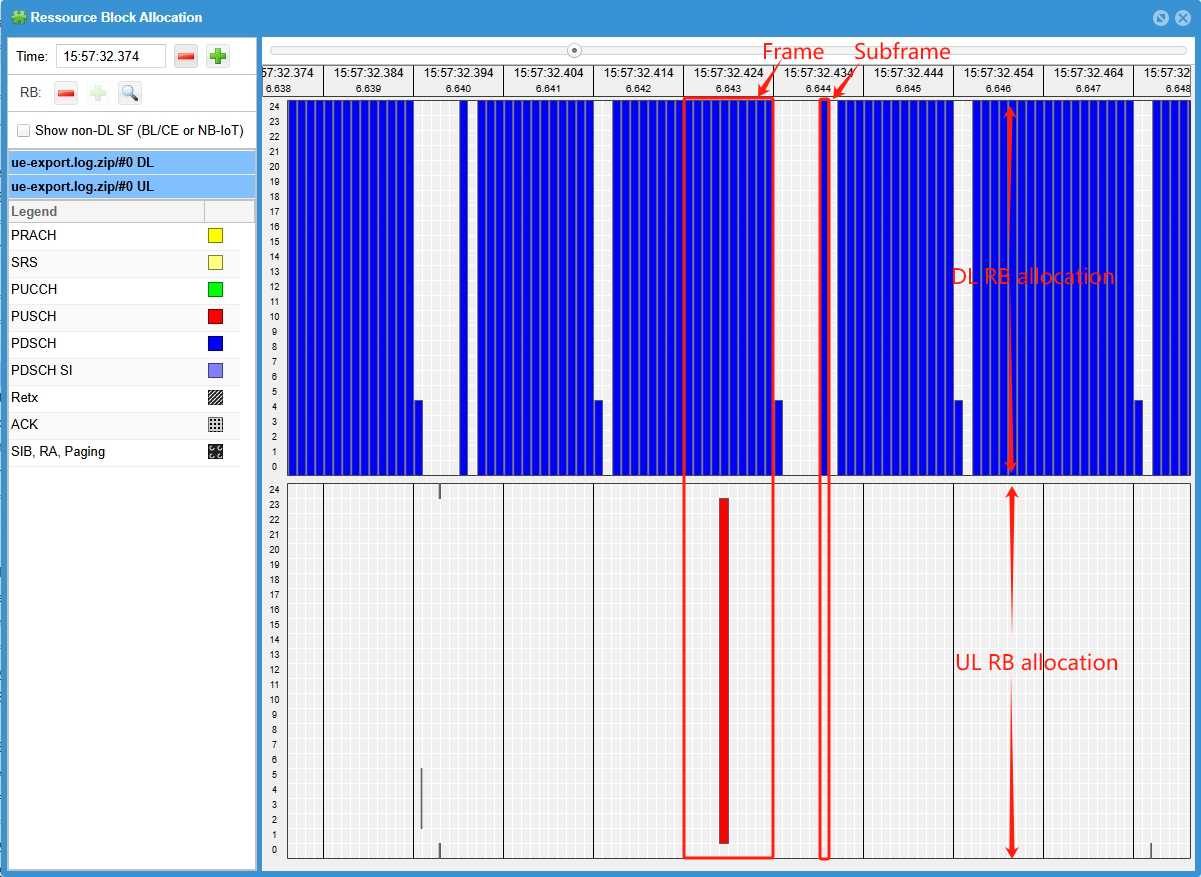}
    \caption{Example of RB Allocation During Testing.}
    \label{fig_rb}
\end{figure}

\subsection{Testing Procedures}


At the beginning of the experiments, the SDN-based NTN gNB and the 5G CN are powered on to establish the NG Application Protocol (NGAP) Stream Control Transmission Protocol (SCTP) connection between them. Once this connection is set up, the NTN cell is activated. Simultaneously, the NTN UE emulator is powered on to initiate the cell search procedure, during which it detects and decodes various signaling information about the NTN cell, such as the cell ID, Master Information Blocks (MIBs), and System Information Blocks (SIBs), as illustrated in Fig.~\ref{fig_uelog}. It is worth noting that System Information Block 19 (SIB19), introduced by 3GPP in Release 17, provides NTN-specific parameters for the serving cell and/or neighboring cells~\cite{TS_38_331}. The left panel of Fig.~\ref{fig_uelog} shows the detailed signaling processing captured at UE. The MIB and SIB messages are logged, as well as the signaling interactions during the UE registration and the establishment of PDU session. The NTN logs such as the TA common, TA UE, elevation and azimuth are also presented. As shown in the right panel of Fig.~\ref{fig_uelog}, SIB19 contains essential information for UE access to NTN systems, including Timing Advance (TA) parameters, ephemeris data, the validity duration of the uplink (UL) synchronization epoch time, and so on. After acquiring the aforementioned information, the NTN UE emulator continuously calculates necessary parameters dedicated for NTN transmission, such as TA common, TA UE, elevation, azimuth, etc. According to TS 38.211~\cite{TR_38_211}, the initial timing advance (TA) is expressed as
\begin{equation}
T_\mathrm{TA} = \left(N_\mathrm{TA} + N_\mathrm{TA, offset} + N_\mathrm{TA, adj}^\mathrm{common} + N_\mathrm{TA, adj}^\mathrm{UE}\right) T_\mathrm{c},
\end{equation}
where $T_\mathrm{c} \approx 0.509 \,\text{ns}$ denotes the fundamental time unit. The parameters $N_\mathrm{TA}$ and $N_\mathrm{TA, offset}$ are identical to those in terrestrial 5G, while $N_\mathrm{TA, adj}^\mathrm{common}$ and $N_\mathrm{TA, adj}^\mathrm{UE}$ are newly introduced for NTN. Specifically, $N_\mathrm{TA, adj}^\mathrm{common}$ compensates feeder link timing and is broadcast from the gNB to the UE through SIB19, as defined in TS 38.213~\cite{TS_38_213}, whereas $N_\mathrm{TA, adj}^\mathrm{UE}$ compensates service link timing and is computed by the UE based on ephemeris information in SIB19. The overall compensation is carried out automatically at both the gNB and the UE. 
\par

Base on the parameters calculated above, the NTN UE emulator changes the timing advance slot for uplink transmission, and then initiates the random access procedure, which may require multiple attempts and fine-tuning. This is due to the presence of multiple frequency converters in the forward and reverse satellite links operating in the Ku band, each potentially introducing additional frequency errors due to hardware imperfections.
\par
Once the random access process is completed, the NTN UE emulator proceeds with the Protocol Data Unit (PDU) session establishment. Upon successful setup, the NTN UE emulator is assigned with an IP address, enabling it to exchange traffic with the 5G CN.
\par
Due to the inherent long propagation delay in the satellite link, the HARQ mechanism is disabled to simplify the testing process.
\par

\subsection{Results and Analysis}
\label{subsec_results}

To evaluate and analyze the performance of the NR NTN system over a real GEO satellite link, we conducted experiments focusing on downlink throughput and round-trip time (RTT) performance, as well as the link quality evaluation.
\par

\begin{figure}[!t]
    \centering
    \begin{subfigure}{0.49\textwidth}
        \centering
        \includegraphics[width=0.85\textwidth]{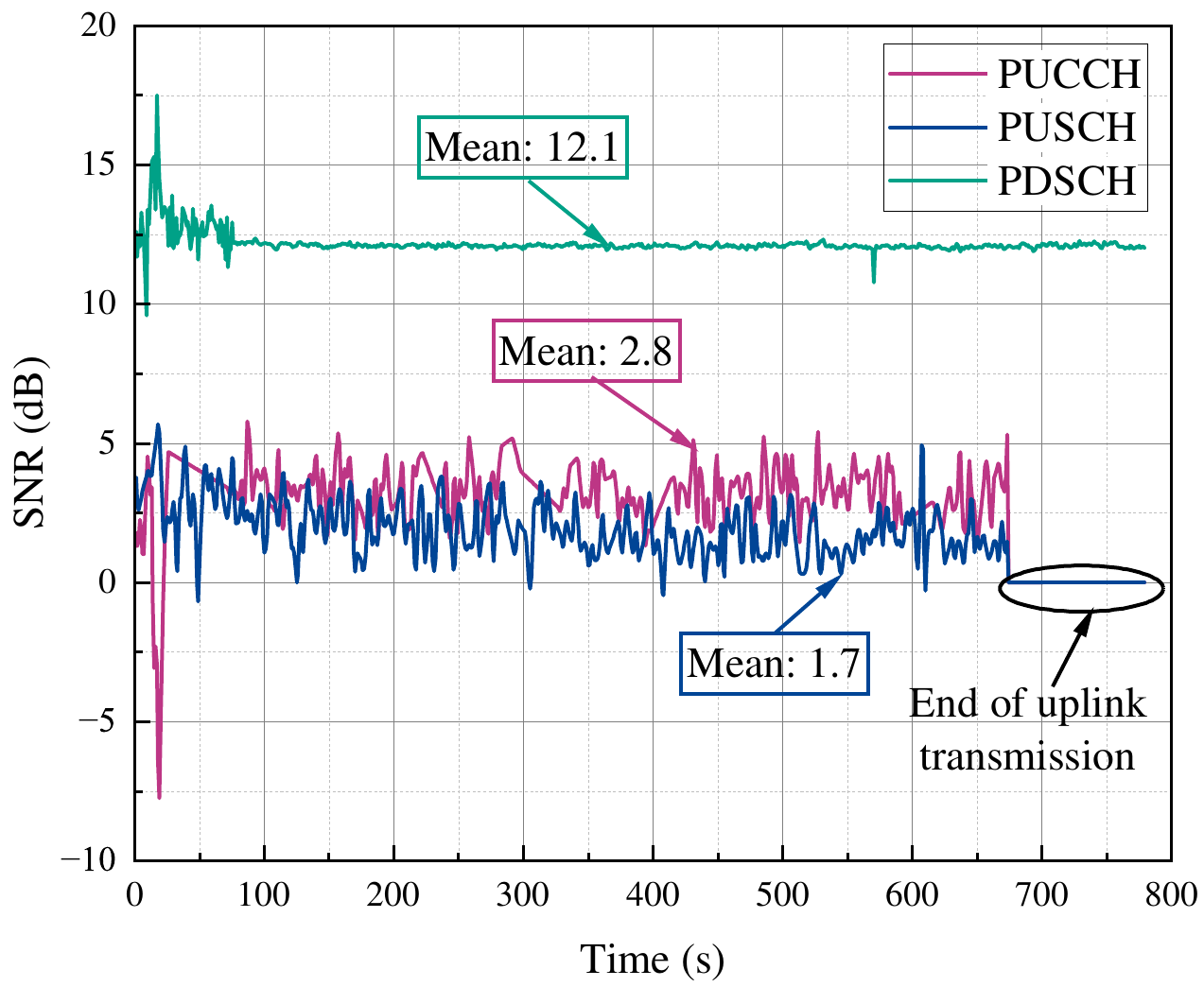}
        \caption{SNR.}
        \label{fig_res_snr}
    \end{subfigure}
    \hfill
    \begin{subfigure}{0.49\textwidth}
        \centering
        \includegraphics[width=0.85\textwidth]{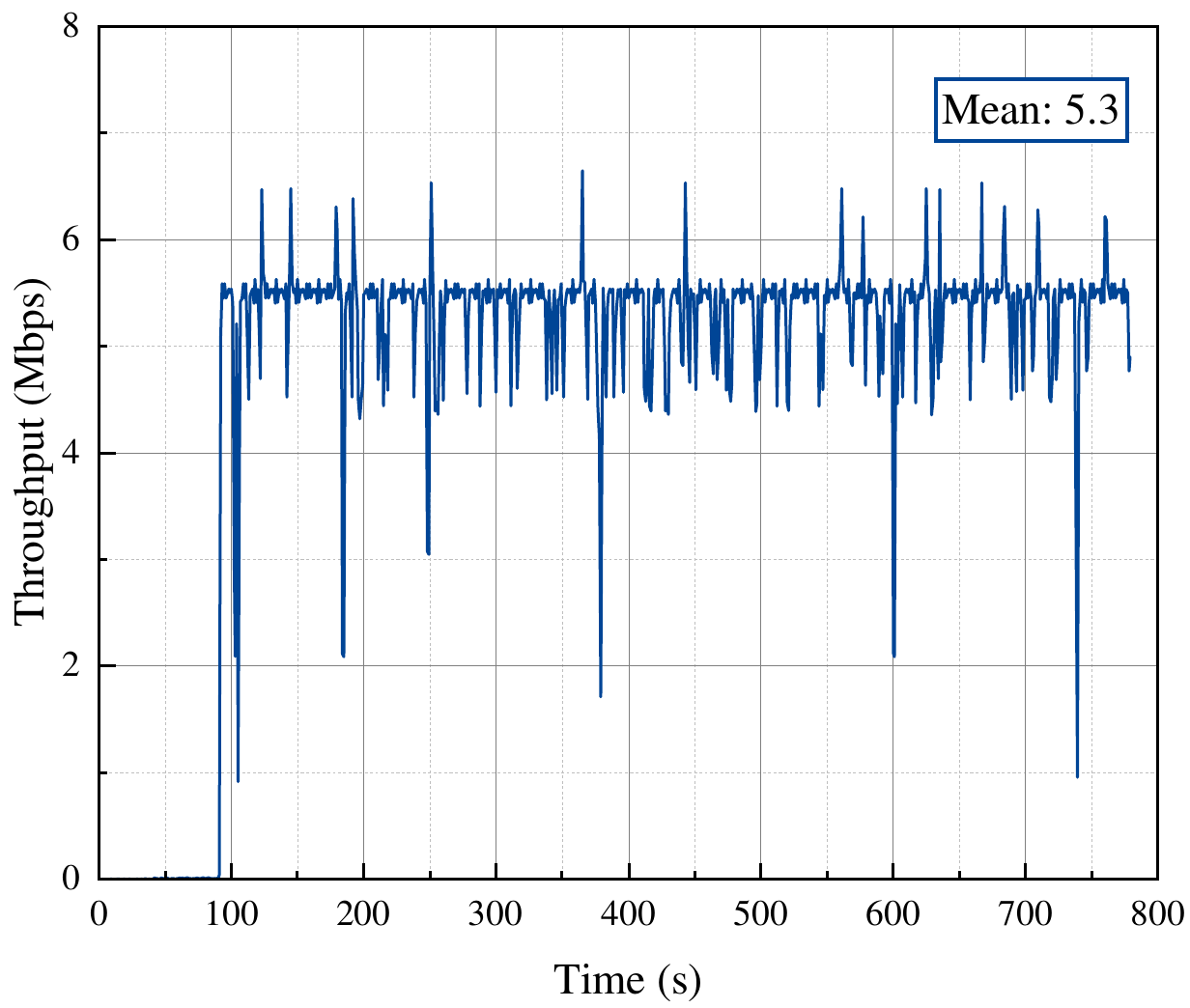}
        \caption{Throughput.}
        \label{fig_throughput}
    \end{subfigure}
    \hfill
    \begin{subfigure}{0.49\textwidth}
        \centering
        \includegraphics[width=0.85\textwidth]{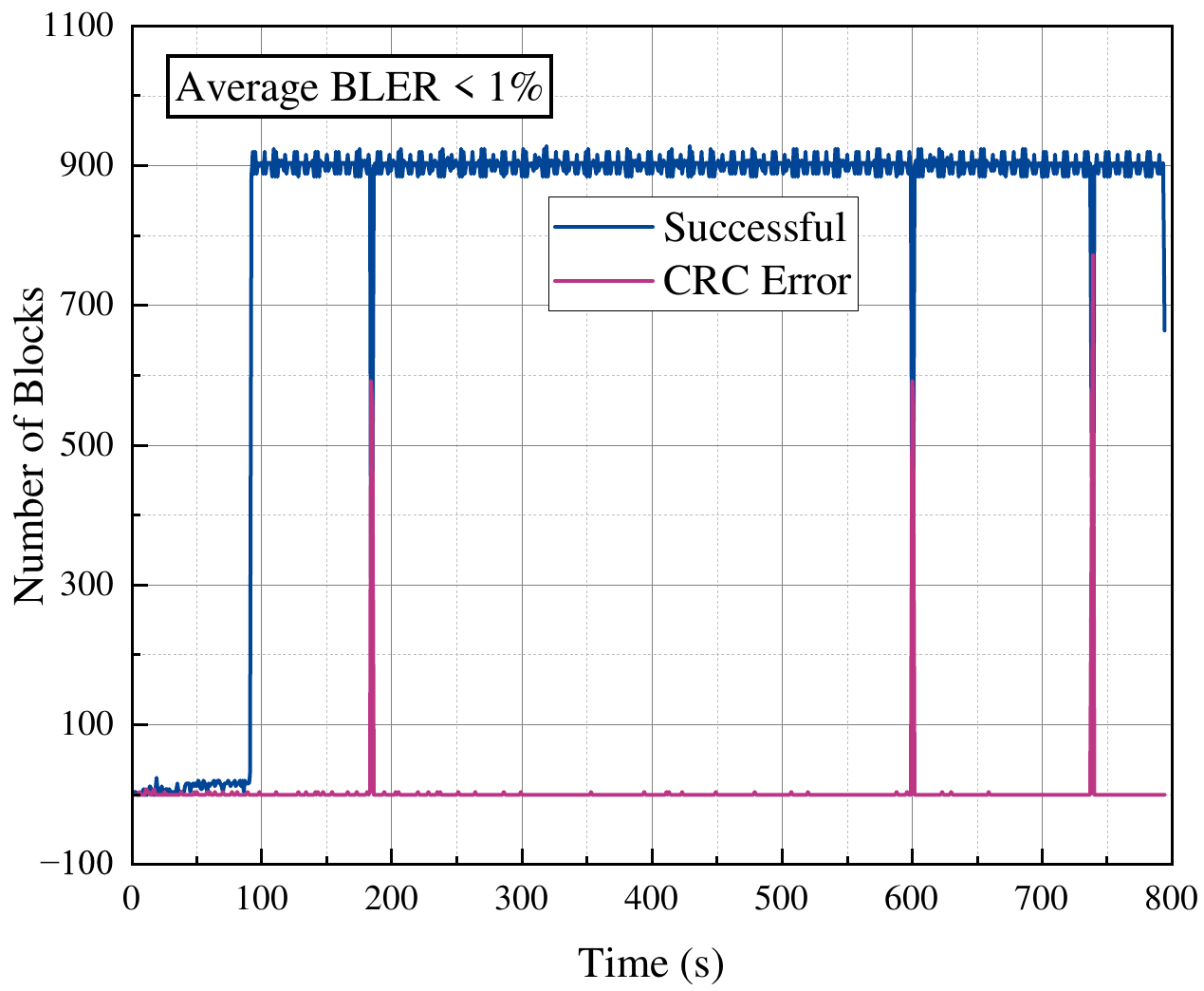}
        \caption{BLER.}
        \label{fig_res_bler}
    \end{subfigure}
    \caption{Performance results of downlink transmission in field experiments.}
    \label{fig_results_cq}
\end{figure}

For the downlink throughput evaluation, we employed the iPerf\footnote{https://iperf.fr/} tool at the NTN gNB to generate full-buffer data traffic transmitted to the NTN UE emulator. For illustration purposes, an example of Resource Block (RB) allocation over ten system frames observed during the test is presented in Fig.~\ref{fig_rb}. We first present the link quality results to validate the correctness of the test setup. Figure~\ref{fig_res_snr} depicts the SNR curves of the PDSCH, PUCCH, and PUSCH. As expected, the downlink SNR is significantly higher than the uplink due to the greater transmission power of the gNB compared with the UE. Specifically, the downlink SNR of the PDSCH measured at the NTN UE emulator is approximately \qty{12}{\decibel}, which leads the NTN gNB to select a Modulation and Coding Scheme (MCS) index of about \num{8}. According to the 3GPP specification~\cite{TS_38_214}, this corresponds to QPSK modulation with a coding rate of \num{0.54}. These observations are further corroborated by the downlink throughput results shown in Fig.~\ref{fig_throughput}, obtained over an approximately 10-minute testing period. The throughput predominantly fluctuates between \num{4} and \num{6}~~\myunit{Mbps}, with an average of \qty{5.3}{Mbps}, corresponding to a spectral efficiency of about \qty{1}{bps\per\hertz}. Although this throughput is considerably lower than that of typical terrestrial NR networks, it is consistent with expectations for the NTN test scenario. Furthermore, the throughput experiences notable drops at several intervals, which align with the BLock Error Ratio (BLER) variations shown in Fig.~\ref{fig_res_bler}. Despite these fluctuations, the average BLER remains below \qty{1}{\percent}, thereby demonstrating the reliability of the radio link throughout the experiments.

\par
\begin{figure}[!t]
    \centering
    \includegraphics[width=0.48\textwidth]{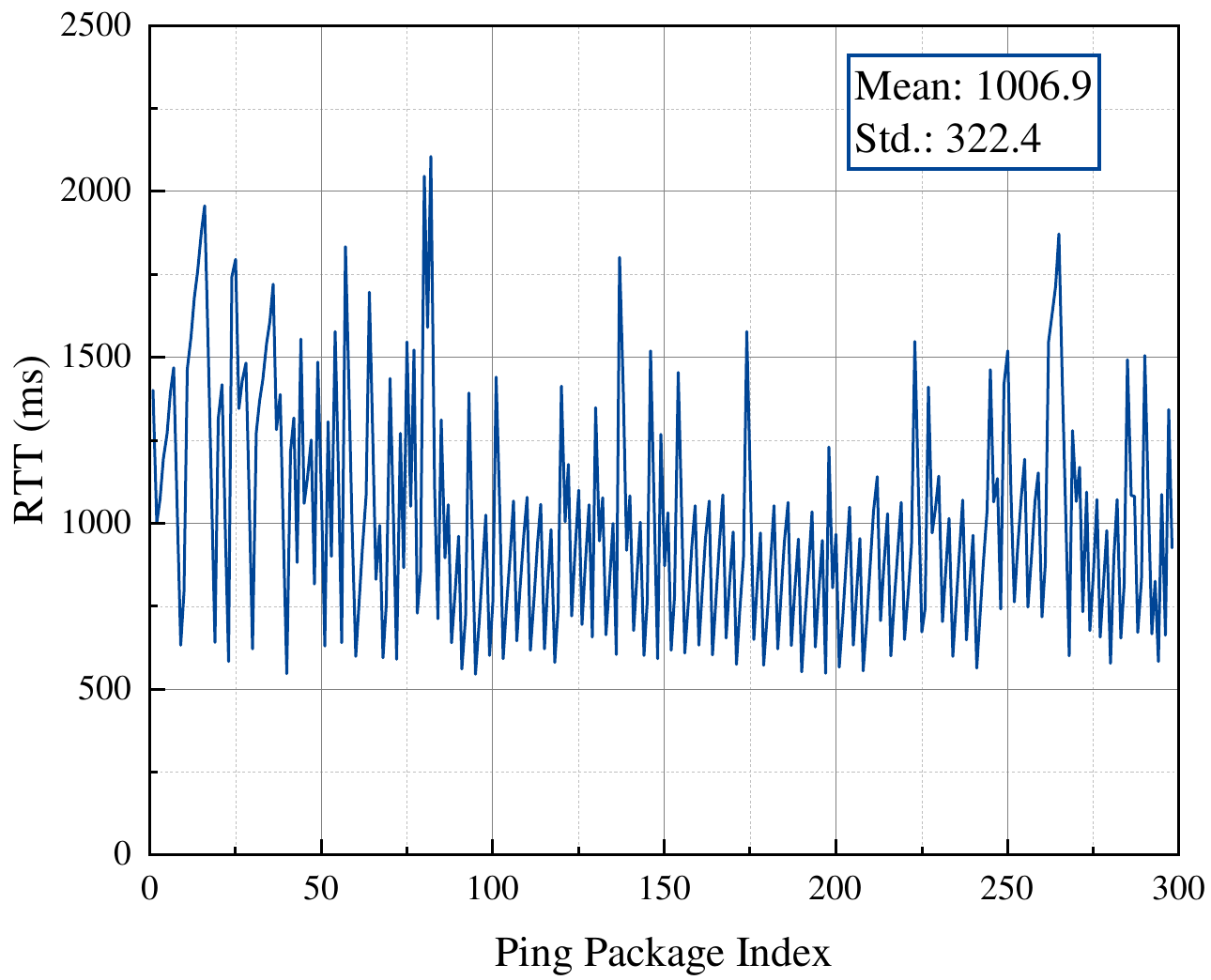}
    \caption{Performance results of RTT in field experiments.}
    \label{fig_rtt}
\end{figure}
For RTT performance assessment, we utilized the ping command. Fig.~\ref{fig_rtt} shows the example results of RTT measurements, whose values ranged from approximately \qty{600}{\milli\second} to \qty{2000}{\milli\second}. Meanwhile, the average RTT is also measured at \qty{1006.9}{\milli\second}, with a standard deviation of approximately \qty{322.4}{\milli\second}. As expected, this exceeds the ideal end-to-end bidirectional propagation delay of a GEO link, which is around \qty{500}{\milli\second}. The significant RTT variation is primarily attributed to uplink transmission limitations. Specifically, the test configuration employed 16 PUSCH HARQ processes without enabling HARQ Mode B~\cite{TS_38_300}, causing uplink transmissions to frequently wait for available HARQ processes, each requiring time equivalent to the end-to-end propagation delay. Furthermore, the substantial number of uplink retransmissions further exacerbated the RTT performance degradation, highlighting the inherent challenges in achieving optimal latency characteristics in NTN systems compared to terrestrial networks. These test results provide valuable insights into the practical performance characteristics of NTN systems operating over GEO satellite links.


\section{Conclusion and Outlook}
\label{sec:Conclusion}

This paper presents a fully functional SDR-based test platform tailored for 5G NTN experimentation, addressing the urgent need for practical validation tools amid evolving 3GPP standards. By utilizing commercial-grade SDR software from Amarisoft and a GEO satellite, we demonstrate the feasibility of establishing an end-to-end 5G NTN communication link between an SDR-based gNB and UE emulator. Our field experiments confirm the platform's capability to support NTN-specific protocol features and operational constraints, including timing advance, extended round-trip delay, and satellite-specific propagation effects. The observed throughput and RTT performance, while constrained by current hardware limitations and satellite link conditions, align with expected characteristics of early-stage NTN systems and reinforce the platform's applicability for real-world testing.
\par
The modularity, reconfigurability, and standards compliance of our platform make it a valuable tool for academic and industrial researchers pursuing performance evaluation, algorithm prototyping, and pre-commercial testing of 5G and beyond-5G NTN technologies. Future work will focus on several key extensions: (i) performance testing under diverse traffic types, network topologies, and antenna configurations; (ii)
enabling transparent mode NTN access and interworking with terrestrial networks; (iii) supporting non-GEO constellations such as LEO and MEO; (iv) integrating new link adaptation and resource management strategies; and (v) evolving the platform toward NTN-enabled 6G use cases. These efforts aim to further bridge the gap between evolving NTN standards and practical deployment readiness.

\section*{Acknowledgment}
%
This work is supported in part by National Natural Science
Foundation of China (NSFC) under Grant No. 62471262,
in part by Zhejiang Provincial Natural Science Foundation
of China under Grant No. LQ23F010012, and in part by
Ningbo University’s Research Start-Up Funding under Grant
No. ZX2023000906.
\bibliographystyle{unsrt}
\bibliography{IEEEabrv,Bib/Ref}
%
%


\end{document}